\documentclass[sigconf, screen]{acmart}

\usepackage{color}
\usepackage{listings}
\usepackage[pdf]{graphviz}
\definecolor{GrayCodeBlock}{RGB}{241,241,241}
\definecolor{BlackText}{RGB}{110,107,94}
\definecolor{BlueTypename}{RGB}{17,86,182}
\definecolor{RedTypename}{RGB}{182,86,17}
\definecolor{GreenString}{RGB}{96,172,57}
\definecolor{PurpleKeyword}{RGB}{184,84,212}
\definecolor{GrayComment}{RGB}{170,170,170}
\definecolor{GrayNumber}{RGB}{200,200,200}
\definecolor{GoldDocumentation}{RGB}{180,165,45}
\definecolor{rhodamine}{RGB}{252,88,170}
\lstdefinelanguage{rust}
{
    columns=fullflexible,
    keepspaces=true,
    frame=single,
    framesep=0pt,
    framerule=0pt,
    framexleftmargin=4pt,
    framexrightmargin=4pt,
    framextopmargin=5pt,
    framexbottommargin=3pt,
    xleftmargin=4pt,
    xrightmargin=4pt,
    basicstyle=\small\ttfamily\color{BlackText},
    keywords={
        true,false,
        unsafe,async,await,move,
        use,pub,crate,super,self,mod,
        struct,enum,fn,const,static,let,mut,ref,type,impl,dyn,trait,where,as,
        break,continue,if,else,while,for,loop,match,return,yield,in
    },
    keywordstyle=\color{PurpleKeyword},
    ndkeywords={
        bool,u8,u16,u32,u64,u128,i8,i16,i32,i64,i128,char,str,
        Self,Option,Some,None,Result,Ok,Err,String,Box,Vec,Rc,Arc,Cell,RefCell,HashMap,BTreeMap,
        macro_rules},
    ndkeywordstyle=\color{RedTypename},
    comment=[l][\color{GrayComment}\slshape]{//},
    morecomment=[s][\color{GrayComment}\slshape]{/*}{*/},
    morecomment=[l][\color{GoldDocumentation}\slshape]{///},
    morecomment=[s][\color{GoldDocumentation}\slshape]{/*!}{*/},
    morecomment=[l][\color{GoldDocumentation}\slshape]{//!},
    morecomment=[s][\color{RedTypename}]{\#![}{]},
    morecomment=[s][\color{RedTypename}]{\#[}{]},
    stringstyle=\color{GreenString},
    string=[b]",
    classoffset=1, 
    alsoletter={-,>},
    keywords={anti_join,batch,cross_join,demux,dest_sink,dest_sink_serde,difference,filter, filter_map,flat_map,flatten,fold,for_each,group_by,identity,inspect,join,map,merge,next_stratum, next_tick,null,persist,reduce,repeat_iter,sort,sort_by,source_iter,source_stdin,source_stream,source_stream_serde,tee,unique,unzip,->,hydroflow_syntax},
    keywordstyle={\color{BlueTypename}\bfseries},
    classoffset=0
}

\newif\ifcomments
\ifcomments
    \providecommand{\shadaj}[1]{{\protect\color{brown}{\bf [shadaj: #1]}}}
    \providecommand{\conor}[1]{{\protect\color{red}{\bf [conor: #1]}}}
    \providecommand{\alvin}[1]{{\protect\color{purple}{\bf [alvin: #1]}}}
    \providecommand{\mae}[1]{{\protect\color{blue}{\bf [mae: #1]}}}
    \providecommand{\joe}[1]{{\protect\color{teal}{\bf [joe: #1]}}}
    \providecommand{\jmh}[1]{{\protect\color{teal}{\bf [joe: #1]}}}
    \providecommand{\david}[1]{{\protect\color{green}{\bf [david: #1]}}}
    \providecommand{\chris}[1]{{\protect\color{violet}{\bf [chris: #1]}}}
    \providecommand{\davidmwei}[1]{{\protect\color{pink}{\bf [david wei: #1]}}}
    \providecommand{\kaushik}[1]{{\protect\color{orange}{\bf [kaushik: #1]}}}
    \providecommand{\justin}[1]{{\protect\color{green}{\bf [justin: #1]}}}
    \providecommand{\mingwei}[1]{{\protect\color{rhodamine}{\bf [mingwei: #1]}}}
    \providecommand{\rithvik}[1]{{\protect\color{red}{\bf [rithvik: #1]}}}
    \providecommand{\nc}[1]{{\protect\color{pink}{\bf [nc: #1]}}}
     \providecommand{\accheng}[1]{{\protect\color{olive}{\bf [accheng: #1]}}}
    \providecommand{\dan}[1]{{\protect\color{purple}{\bf [dan: #1]}}}
  \else
    \providecommand{\shadaj}[1]{}
    \providecommand{\conor}[1]{}
    \providecommand{\alvin}[1]{}
    \providecommand{\mae}[1]{}
    \providecommand{\joe}[1]{}
    \providecommand{\jmh}[1]{}
    \providecommand{\david}[1]{}
    \providecommand{\chris}[1]{}
    \providecommand{\davidmwei}[1]{}
    \providecommand{\kaushik}[1]{}
    \providecommand{\justin}[1]{}
    \providecommand{\mingwei}[1]{}
    \providecommand{\rithvik}[1]{}
    \providecommand{\nc}[1]{}
    \providecommand{\accheng}[1]{}
    \providecommand{\dan}[1]{}
\fi

\AtBeginDocument{%
  \providecommand\BibTeX{{%
    \normalfont B\kern-0.5em{\scshape i\kern-0.25em b}\kern-0.8em\TeX}}}

\copyrightyear{2023}
\acmYear{2023}
\setcopyright{rightsretained}
\acmConference[ApPLIED 2023]{The 5th workshop on Advanced tools, programming languages, and PLatforms for Implementing and Evaluating algorithms for Distributed systems}{June 19, 2023}{Orlando, FL, USA}
\acmBooktitle{The 5th workshop on Advanced tools, programming languages, and PLatforms for Implementing and Evaluating algorithms for Distributed systems (ApPLIED 2023), June 19, 2023, Orlando, FL, USA}
\acmDOI{10.1145/3584684.3597272}
\acmISBN{979-8-4007-0128-3/23/06}

\begin{document}

\title[Invited Paper: Initial Steps Toward a Compiler for Distributed Programs]{Invited Paper: Initial Steps Toward a Compiler for \linebreak Distributed Programs}

\author{Joseph M. Hellerstein}
\email{hellerstein@berkeley.edu}
\orcid{0000-0002-7712-4306}
\affiliation{%
  \institution{UC Berkeley}
  \streetaddress{387 Soda Hall \#1776}
  \city{Berkeley}
  \state{CA}
  \country{USA}
  \postcode{94720-1776}
}

\author{Shadaj Laddad}
\email{shadaj@berkeley.edu}
\orcid{0000-0002-6658-6548}
\affiliation{%
  \institution{UC Berkeley}
  \streetaddress{387 Soda Hall \#1776}
  \city{Berkeley}
  \state{CA}
  \country{USA}
  \postcode{94720-1776}
}

\author{Mae Milano}
\email{mpmilano@berkeley.edu}
\orcid{0000-0003-3126-7771}
\affiliation{%
  \institution{UC Berkeley}
  \streetaddress{387 Soda Hall \#1776}
  \city{Berkeley}
  \state{CA}
  \country{USA}
  \postcode{94720-1776}
}

\author{Conor Power}
\email{conorpower@berkeley.edu}
\orcid{0000-0002-0660-5110}
\affiliation{%
  \institution{UC Berkeley}
  \streetaddress{387 Soda Hall \#1776}
  \city{Berkeley}
  \state{CA}
  \country{USA}
  \postcode{94720-1776}
}

\author{Mingwei Samuel}
\email{mingwei@shv.com}
\orcid{0009-0004-9873-6266}
\affiliation{%
  \institution{Sutter Hill Ventures}
  \streetaddress{Bldg A 200, 755 Page Mill Rd}
  \city{Palo Alto}
  \state{CA}
  \country{USA}
  \postcode{94304}
}

\renewcommand{\shortauthors}{Hellerstein, et al.}

\begin{abstract}
  In the Hydro project we are designing a compiler toolkit that can optimize for the concerns 
  of distributed systems,
  including scale-up and scale-down, availability, and consistency of outcomes across replicas. 
  This invited paper overviews the project, and provides an early walk-through of the kind of optimization that is possible. We illustrate 
  how type transformations as well as local program transformations can combine, step by step, to convert a single-node program
   into a variety of distributed design points that offer the same semantics with different performance and deployment characteristics.
\end{abstract}

\keywords{distributed computing, programming languages, compiler, query optimization, dataflow}



\maketitle

\section{Introduction}
An ongoing thread in distributed computing is the development of 
new programming models and languages built on formal models that can
simplify the challenges developers face in building distributed software. This thread includes programming languages like Dedalus~\cite{dedalus}, Bloom~\cite{bloom,blooml}, LVars~\cite{lvars}, Lasp~\cite{lasp}, Datafun~\cite{datafun} and Gallifrey~\cite{gallifrey}, as well as data structures like CRDTs~\cite{shapiro2011comprehensive} and their realization in libraries like Automerge~\cite{automerge}.

This short paper is part of an evolution from language design to a full stack for distributed programming.
Is it possible to build a language stack (multiple surface languages with shared compilation, debugging and deployment) that addresses the concerns of developers writing distributed programs? How much work can be done automatically? Of what remains, what is amenable to compiler assistance and human review?  Can compiled code compete with hand-written code? Can a compiler discover optimizations that humans do not?
This paper is an early snapshot of our work in this domain. It does not claim to answer all of these questions, nor even to answer any one of them definitively. We narrow our focus here to automatic compiler transformations, with much of the discussion driven from simple examples. In short, this paper is intended as a progress report and an opening for community engagement.

Traditional optimizing compilers concern themselves with efficient use of computing resources including the various aspects of CPUs, GPUs, memory and interconnects. All of these concerns exist in distributed systems of course, but are augmented by concerns that are endemic to distributed systems: notably communication, partitioning of work across machines, fault tolerance,  concurrency and consistency of data and outcomes,
and respect for invariants related to security, privacy and governance. We assume that traditional optimizer toolkits like LLVM~\cite{llvm} will continue to serve the purposes of optimizing the code that runs on each machine; we focus on the challenge of providing abstractions and compilers for the unique distributed aspects of modern programs.

\begin{figure}[t]
    \centering
    \includegraphics[width=3in]{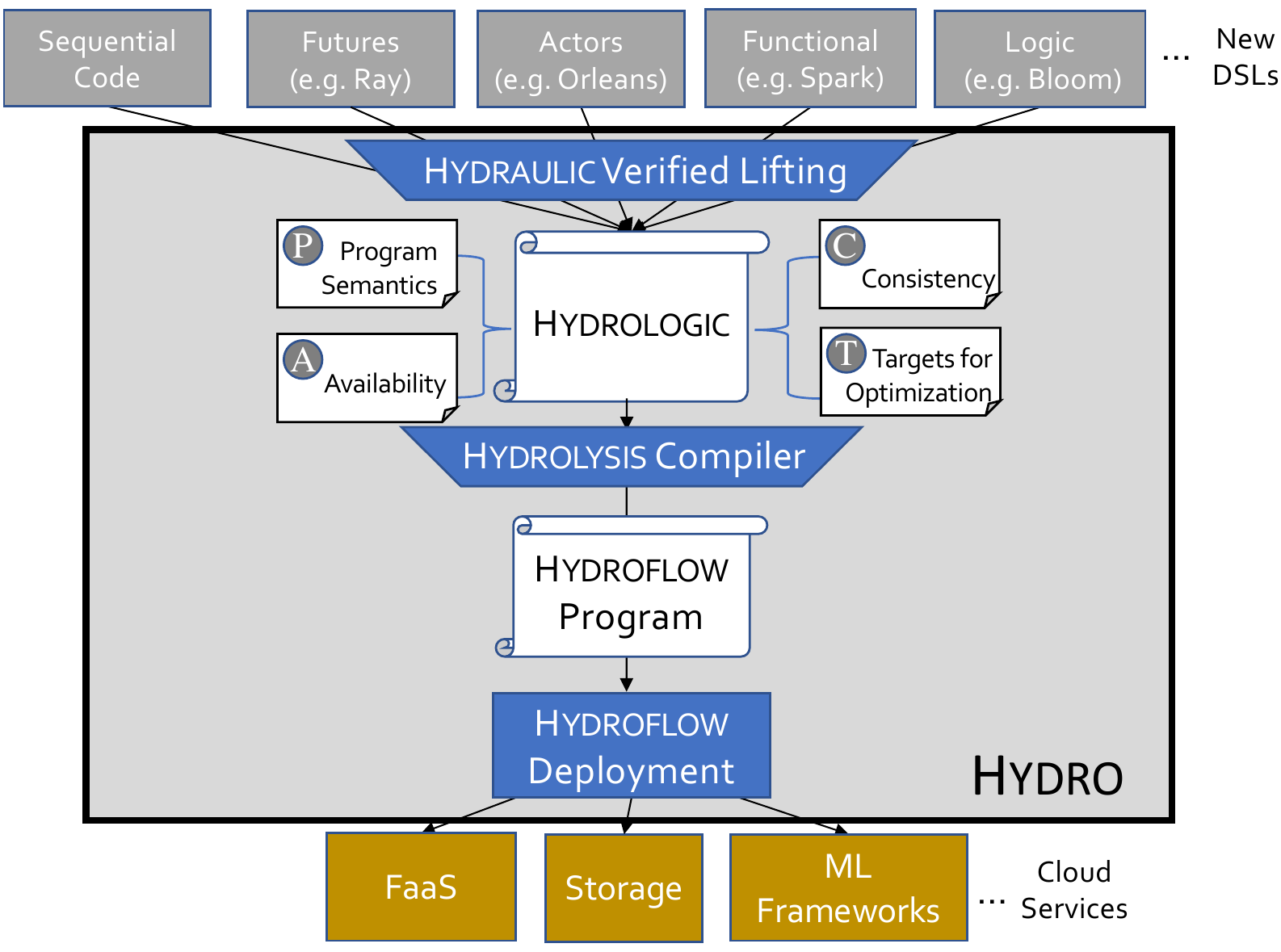}
    \caption{The Hydro compiler stack~\cite{hydro}.}
    \label{fig:hydro}
\end{figure}

We ground our discussion in the context of the Hydro project, a multi-year effort at UC Berkeley and Sutter Hill Ventures. We laid out our vision for Hydro in an earlier paper~\cite{hydro}, proposing a compiler stack (Figure~\ref{fig:hydro}) with multiple components. 
As the highest level of input to Hydro, we hope to support a multitude of distributed programming styles, much like LLVM provides a compiler stack 
for a multitude of sequential programming languages. Also like LLVM, Hydro envisions multiple layers of intermediate representation languages (IRs) that can serve as common ground for program checks and transformations, providing developers with various entry points to work deeper in the stack as they see fit. The top layer of Hydro we call \emph{Hydraulic}: a system to lift low-level code from legacy interfaces into a higher-level declarative distributed IR we dub \emph{Hydrologic}. The next layer down is an optimizing compiler we call \emph{Hydrolysis}, which takes Hydrologic and compiles it to run on multiple instances of a single-threaded, asynchronous dataflow IR called \emph{Hydroflow}. Work toward Hydraulic and Hydrologic is ongoing, with initial results beginning to emerge~\cite{katara}. 

The focus for this paper is on Hydrolysis---specifically we want to explore the feasibility of a compiler that can produce Hydroflow programs optimized for different deployment objectives.
We consider the potential for a \emph{transformation-based} compiler that can correctly modify a Hydroflow program, step-by-step, into one or more alternative Hydroflow programs that offer the same results with different performance or deployment characteristics. Like many dataflow systems, Hydroflow is an extension of the relational algebra. As such it is amenable to 
the kinds of query optimizations pioneered for database systems, as well as a host of optimizations that are apropos for general-purpose distributed programs.  Following the model of the widely-used Cascades query optimizer~\cite{cascades} 
and the renewed enthusiasm in the programming languages community for e-graphs~\cite{egraphs,egg},
we view query optimization as a problem of \emph{program transformation}. Basically, we assume we can translate from Hydrologic to \emph{some} semantically-equivalent \emph{single-node} Hydroflow program (``execution plan'') in a naive fashion---essentially via parsing Hydrologic without concern for distributed deployment or performance.
The compiler's job then is to search the space of semantically equivalent dataflow programs running on one or more machines, and choose the
configuration that is most desirable according to an objective function on various metrics (e.g. cloud costs, runtime performance, resilience to failures, etc.) with various constraints or invariants on how data and communication are to be managed.

\mingwei{Not sure how important the exact technique is, but here we'd be using e-graphs instead}\jmh{Gotcha.} In the style of Cascades and egraphs, we envision optimization via repeated application of simple local program transformations, using compact memoization to keep track of prior states and avoid repeated work. The basic loop is this: the current Hydroflow program is stored in a memoization structure, and a local \emph{transformation rule} (a kind of peephole optimization) is applied to a small segment of the Hydroflow program to get another semantically equivalent Hydroflow program that we have not seen before. In the presence of a cost model and objective function, pruning is applied to the memo structure and search space to keep only those candidates that may participate in an optimal outcome.
This process repeats until all possible semantically equivalent dataflows have been generated or ruled out as suboptimal via pruning. 

This paper does not deliver on the full vision of the Hydrolysis optimizer, though we sketch next steps in Section~\ref{sec:future}. Our discussion here is an early demonstration of manually applied transformation rules that  achieve useful optimizations for correctly distributing programs across machines. 
We do not pretend to offer a comprehensive set of such transformations as of yet.  Our goal is to document our growing confidence in the potential for an optimizing compiler---specifically one based in a dataflow model---to meaningfully assist in the development of efficient, correct distributed programs.

\subsection{Why Dataflow?}
One of the signatures of the Hydro project is the opinionated decision to use a high-performance dataflow kernel as its lowest-level language. This may not be an obvious choice to researchers in classical distributed systems.

At the outset, we were confident about the scalability of a dataflow IR because of the success of prior dataflow engines at auto-parallel\-ization. Languages like SQL and Spark have put effortless scaling into the hands of programmers for decades, even as other broad efforts at parallel and distributed programming languages failed~\cite{cccblog}. The runtimes for those data-centric languages are parallelized dataflow engines. Unlike Spark and Hadoop, dataflow runtimes for SQL have targeted heterogeneous workloads and performance goals, including low-latency infrastructure. There was reason to be confident that dataflow can meet our low-level performance goals \emph{and} scale with ease, as we discuss below.

More generally, by using dataflow we gained access to a long tradition of database and compiler literature on optimization. The database literature is founded on the duality between dataflow algebra and high-level query languages like SQL---i.e., Codd's Theorem~\cite{codd1970relational}, the basis of his Turing Award. Because Hydroflow is so close to query languages like SQL or Datalog, we can apply the full body of database theory and practice to our compiler runtime. 

One of the benefits of dataflow and query languages that we exploit is the ease of refactoring code. Auto-parallelization of sequential code involves teasing apart a monolithic program into separable components, and dataflow makes this almost trivial. Every dataflow program is a graph of producers and consumers, so refactoring a program into separate software components is almost as simple as changing local dataflow pipes into network channels. Of course this requires care to maintain program semantics, as we'll discuss below. 
By contrast, as any software engineer knows, it can be very hard to refactor a sequential program without breaking it---this is especially true of complex programs like the Paxos variants we have been building. 

Second, the primary syntactic feature of a dataflow language is the explicit \emph{specification of data dependencies}. Our transformations use data dependencies to analyze the interplay between components, and reason about the implications of placing components on separate nodes across networks. 
By contrast, data dependencies in sequential programs are implicit, based on complex \emph{program slicing}~\cite{slicing} that has to account for issues of control flow and mutable state that are absent in dataflow models. 

In addition to these overarching benefits, database theory provides us with additional technical tools that are relevant to distributed systems.
One of particular interest is the ability to use simple checks for \emph{monotonicity}, the property 
that the CALM Theorem shows to be both necessary and sufficient for consistent results in the absence of coordination~\cite{calm}. We can exploit this to decouple code freely across multiple machines without thought for ordering, synchronization or coordination. Another feature of interest is the availability of functional dependencies to describe state invariants that ensure safe partitioning (sharding) of code and state. A third tradition is the body of literature on data provenance~\cite{cheney2009provenance}, which allows data dependencies to be analyzed in subtle ways, with applications to distributed systems including use in efficient fault injection~\cite{blazes}. 
These topics are beyond the scope of this paper.

\subsection{Hydroflow and Prior Work}
The open-source implementation of Hydroflow~\cite{hydroflow} provides the concrete setting for our discussion in this paper. As input, the Hydroflow system takes single-node Hydroflow specs embedded in Rust programs, which can use networking components to communicate with each other. Hydroflow provides the libraries, support routines and compilation scaffolding to allow the Rust compiler (which uses LLVM) to emit high-performance code on each individual node. The details of Hydroflow can be found in the online Hydroflow book~\cite{hydroflow-book}; we provide a brief overview here. 

At a high level, Hydroflow is similar to many dataflow runtimes and languages, ranging from database system internals going back to System~R~\cite{systemr}, Ingres~\cite{ingres}, and later extensible runtimes like Volcano~\cite{graefe1994volcano}. Modern readers may be more familiar with contemporary data analytics libraries like Spark~\cite{spark}, Timely Dataflow~\cite{timely} and Pandas~\cite{pandas}. Hydroflow targets somewhat different performance and correctness goals than the prior work:

\vspace{0.5em}
\noindent\textbf{Machine Model}. As a low-level IR, the goal for Hydroflow is to support programs that can be distributed across both cores and machines at any scale from a single box to the globe and beyond~\cite{anna}. Hydroflow models the behavior of a set of independent communicating agents (``nodes''), each with its own local state and logical clock. 
Hydroflow assumes only point-to-point communication, with no assumptions of reliability or ordering on channels, nor built-in facilities for broadcast.
In practice, a sender can communicate only with a receiver for whom it has an address in its local state.
This captures a standard asynchronous model in which messages between correct nodes can be arbitrarily delayed and reordered, and formally all messages are eventually delivered after an infinite amount of time~\cite{dwork1988consensus}, but in practice delays can be managed via timeouts, which can be specified to arrive as external stimuli to the system. The Hydroflow runtime and language make no further assumptions about failures of nodes or message delivery. 
The runtime offers general MPMD 
setups where
each node can have different programs and data; more uniform setups are possible as well.
Hydroflow does assume a globally-defined namespace for network endpoints (e.g. \texttt{IP:port} for internet deployments), but it does not assume individual nodes have knowledge of node membership. 
The runtime assumes no built-in mechanism for shared state across nodes in the language, but shared memory queues are supported transparently as a communication channel when feasible.

\vspace{0.5em}
\noindent\textbf{A Single-Node Kernel With Networking Support}. Many modern dataflow systems from the ``Big Data'' era are designed for parallel execution across multiple nodes. In the Hydro stack, any cross-thread ``global'' model---be it for analytics, live services, or other applications---is the purview of the higher-level Hydrologic language, which is compiled down into Hydroflow. Hydroflow itself is a single-threaded language intended to be run on a single core, with communication support (both shared-memory and networking) allowing multiple Hydroflow instances to run in parallel and communicate efficiently. Using a rough analogy to parallel database systems, Hydrologic's global view is akin to SQL, whereas Hydroflow is a ``query plan'' language and compiler for an individual core participating in the execution of a parallel query.

\vspace{0.5em}
\noindent\textbf{Low-latency Data Handling}. 
Many Big Data and Warehousing-centric systems focus on throughput and bulk-synchronous processing. While this is possible in Hydro, we also target low-latency performance for handling asynchronous network events. In this sense Hydroflow is closer to the Click router~\cite{click} than Big Data systems like Spark. Like Click, Hydroflow includes support for efficiently managing ``push'' and ``pull'' dataflow operators, harnessing the Rust compiler's monomorphization techniques to the task of compiling push/pull dataflows into highly-efficient code that is aggressively ``inlined''~\cite{mingweims}.

\vspace{0.5em}
\noindent\textbf{Algebraic Typing for Distributed Consistency}. Hydroflow builds on research in exploiting formal properties like monotonicity~\cite{calmcacm} for assessing the distributed consistency properties of programs. Like Bloom~\cite{bloom}, it provides a rich dataflow model for composing complex programs, and non-monotonicity analysis to identify program locations that require coordination for consistency. Like LVars~\cite{lvars}, $\mbox{Bloom}^L$, Lasp~\cite{lasp}, Gallifrey~\cite{gallifrey} and Datafun~\cite{datafun}, it uses algebraic properties of join semi-lattices---namely Associativity, Commutativity and Idempotence (a.k.a ``ACID 2.0''~\cite{quicksand})---to distinguish monotonic code fragments from those that require coordination for consistency. Hydroflow is unique in formally modeling the properties of the dataflow runtime itself using join semi-lattices. 

Hydroflow's modeling of dataflow as a join semi-lattice drives a number of our optimization examples below, so we discuss it in more detail in the next section.

Before proceeding, we should address common concerns about performance.
Empirically, Hydroflow's performance is hitting performance targets we set at the outset of the project. For example, a Hydroflow implementation of Compartmentalized Paxos~\cite{compartmentalized} provides better latency \emph{and} peak throughput than the original handwritten Scala code that was state-of-the-art two years ago~\cite{hydroblog}. Similarly, a Hydroflow implementation of the Anna key-value store outperforms the original handwritten C++ code and matches its linear scaling under conflict~\cite{hydroblog}; the original Anna paper was already providing performance under contention that was orders of magnitude faster than research and production systems like Redis and Masstree~\cite{anna}. Raw performance is no longer one of our primary concerns; optimization is the next challenge.

\begin{figure*}[h]
\begin{minipage}{.45\textwidth}
\begin{lstlisting}[language=rust]
source_stream(shopping) -> [0]lookup_class;
source_iter(client_class) -> [1]lookup_class;
lookup_class = join() 
  -> map(|(client, (li, class))| ((client, class), li))
  -> group_by(Vec::new, Vec::push) 
  -> map(|m| (m, out_addr)) -> dest_sink_serde(out);
\end{lstlisting}
\end{minipage}
\begin{minipage}{.55\textwidth}
\pagebreak
    \centering
    \includegraphics[height=1in]{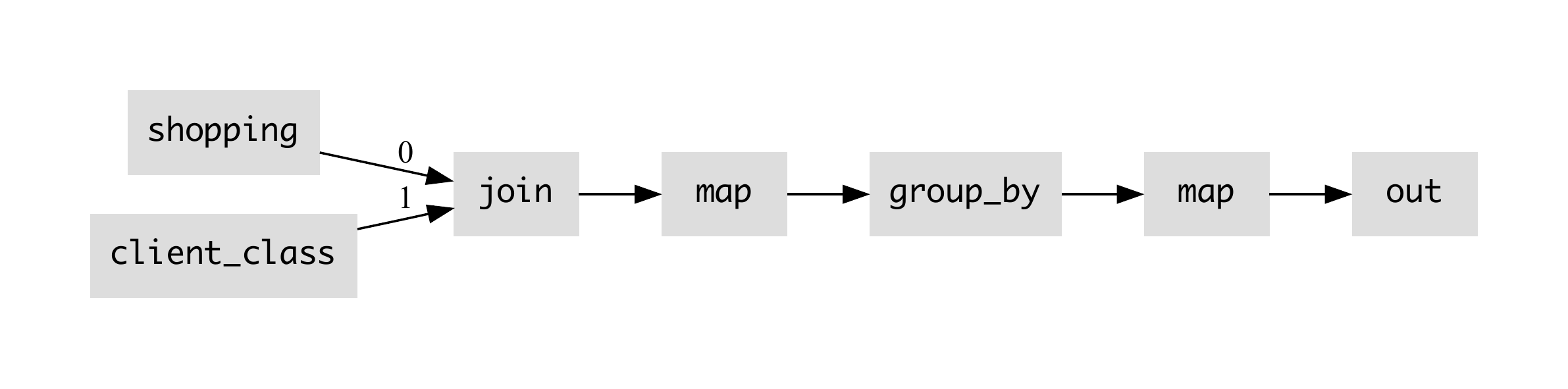}
\end{minipage}
    \caption{Original Flow}
    \label{fig:orig}
\end{figure*}

\section{Dataflow, Networks, and Lattices}
\label{sec:background}
Dataflow is a widely-used programming model for composing simple data operators into complex programs represented as directed graphs. In many dataflow systems, the operators have formal semantics, often as a superset of the relational algebra. By contrast, the semantics of ``edges'' in the dataflow are often implicit in an execution model. Typically, the implicit assumption is that an edge from a producer operator \texttt{P} to consumer operator \texttt{Q} (denoted \texttt{P~->~Q}) indicates that \texttt{P} delivers a \emph{stream} of data to \texttt{Q}. This stream semantics implies an ordering constraint: if \texttt{P} delivers data in a certain order, \texttt{Q} will receive it in that order. In Section~\ref{sec:bp} we discuss how this can be formalized, but for now we can imagine that the producer implicitly assigns a sequence number to each item it sends, and the consumer is guaranteed to receive those items in monotonically increasing order of those sequence numbers.

Dataflow operators can also pass data over a network, a special kind of dataflow edge.
A ``network edge'' of this type \emph{downgrades} the type of communication from ordered streams, by garbling the ordering, batching, and number of transmissions of each item in the stream.
If we again imagine implicit sequence numbers at the producer, the consumer has no guarantee of receiving data over a network edge by monotonically increasing position\footnote{Some networking protocols like TCP offer reliable ordered delivery. These protocols are not a panacaea for many applications, however---it's quite common for TCP sessions to terminate unpredictably. As a result, many long-running services layer their own solutions to ordering and reliability on top of multiple unreliable TCP sessions~\cite{helland2012idempotence}.}. In the absence of these guarantees, it is difficult to reason about the consistency of program outcomes across multiple executions. In particular, if a program with network edges is replicated, the replicas may be inconsistent; alternatively, if such a program crashes and is re-executed (e.g. by a recovery protocol) the second run may not match the prefix of outcomes that happened in the initial run.

To address these concerns,
various projects across decades of work have explored the idea that \emph{certain operators remain consistent even when run on networked edges and/or replicated}. If the operators are inherently Associative, Commutative and Idempotent in their handling of input data, they \emph{upgrade} the dataflow back to consistency of outcomes across networked executions (see~\cite{sagas,quicksand,shapiro2011comprehensive,lvars, lasp, gallifrey, datafun}, etc.) Mathematically, an operator with these properties is a \emph{join semi-lattice}~\cite{shapiro2011comprehensive} (henceforth we will just use the term ``lattice''), and monotonic. 
These monotonic lattice operators will produce identical outcomes in the face of data arriving in different batches (associativity), orders (commutativity) or multiplicity (idempotence). These properties are compositional, so a dataflow composed of join semi-lattices is itself a composite join semi-lattice.
If we can transform our code---whether it be the entire program, or just a component---to use lattices exclusively, we can rest easy about the distinctions between local and networked edges within that scope. The data types will ensure deterministic outcomes across networks\footnote{The CALM Theorem~\cite{calmcacm} proves that this relationship is bidirectional: programs are consistent across unreliable network edges if \emph{and only if} they provide monotonicity guarantees of the form guaranteed by lattices. CALM was proven in a formal framework of distributed logic programming rather than distributed algebra, so there are still some technical details to be done to apply this argument to a language like Hydroflow, but the intuition is fairly direct.}. 
\conor{We kind of use ACI and monotonic interchangeably throughout the text but I think this will be very confusing to a reader. e.g. what does something growing monotonically have to do with tolerance to network batching and duplication? This section would be the best place to bridge these concepts.} \jmh{Better now?}

Given this background, one of the goals of optimization in Hydro is to transform programs to make liberal use of lattice operators. To ensure consistency----that is, determinism across runs and replicas---code segments using non-lattice operators either must run on a single sequential core, or, if distributed, must establish consensus on the order of operations using a protocol like Paxos~\cite{paxos}, which often negatively affects latency and availability~\cite{calmcacm}.

Because lattices lie at the heart of our goals of correctly auto-distributing programs, we focus on foundational lattice-oriented transformations in this paper. There are of course many more transformations that are relevant to distributed program optimization. Two that we have explored extensively in implementing Paxos over Hydroflow are auto-decoupling of subprograms and auto-partitioning (sharding) of code and state~\cite{hydroblog}. However here we stay focused on initial optimizations that demonstrate our ability to safely distribute a simple example in multiple ways.

\section{A Classic Scenario: Shopping Carts}

To illustrate the potential for Hydrolysis, we show how a compiler can take a simple single-node Hydroflow program and transform it step-by-step into a semantically equivalent distributed alternative with a clever twist from the literature. Specifically, we consider the classic problem of implementing a shopping cart, inspired by the Amazon Dynamo paper~\cite{dynamo}. Our goal will be to step through a sequence of individual transformations in the Hydrolysis search space, as an example of the transformation paths that Hydrolysis would enumerate. All the code we show below runs correctly in Hydroflow, and is available in full at \url{https://github.com/hydro-project/hydroflow/tree/applied23/hydroflow/examples/shopping}.

Given a single-node implementation of a shopping cart system, we
partition the program between client and server, replicate the servers for fault tolerance, and introduce an optimization from Conway et al.~\cite{blooml} to work with lattices throughout---hence allowing not only shopping but also checkout to proceed without using any distributed coordination.

Our naive Hydroflow code for the shopping carts is shown in Figure~\ref{fig:orig}, along with an auto-generated dataflow diagram of the code. We envision two classes of shopping data, one for basic customers and one for ``premium'' customers. The type of the shopping data streaming in is of the form \texttt{Stream<ClientLineItem>}---an unbounded list of requests. \texttt{ClientLineItem} is a nested pair \texttt{(client: usize, (item: String, qty: i16))} representing a request from a client with a non-negative integer ID (Rust's \texttt{usize} type) to add a quantity of a specific item to their shopping cart, or delete a quantity of an item (via negative \texttt{qty}).  In this simple program, a shopping cart is similar to string data, but rather than being an ordered list of characters, it is an ordered list of \texttt{ClientLineItems}\footnote{It is tempting to assume that shopping requests would be better represented as a set than a stream. The problem is that the quantity of each item needs to be handled carefully. Imagine that a customer orders 2 apples, then orders 2 more apples, then deletes 4 apples. In the end they truly want 0 apples. Two problems arise. One is that sets are idempotent, but counting/summing is not. So the following two sets are equivalent $\{(apple \times 2),(apple \times 2)\} = \{(apple \times 2)\}$, but the following streams are not equivalent $[(apple \times 2), (apple \times 2)] \ne [(apple \times 2)]$. The second problem is that the semantics of deletion and insertion may not be commutative: in some applications, we may ignore "overdrafts" that go below 0. For example, in some definitions, $[(apple \times -4)], [(apple \times  4)] = [(apple \times 4)]$ because the deletion on an empty cart is ignored. Stream semantics ensure these issues are unambiguous.}.

We begin in Figure~\ref{fig:orig} with a single-node Hydroflow implementation that we envision being generated naively from a distribution-agnostic Hydrologic spec. 
We will walk through this code in some detail; subsequent snippets in the paper use substantially the same operators, just reconfigured via various transformations.

In the Hydroflow language, $\texttt{->}$ represents a stream of data flowing from a producing \emph{operator} to a consuming \emph{operator} on a single node. Hydroflow offers a variety of operators familiar from relational algebra and functional languages like Spark or Pandas, including the ability to embed "user-defined functions" (i.e. arbitrary single-node sequential code) in operators like \texttt{map} and \texttt{reduce}.

Taking the code one line at a time, we begin in Line~1 with a \texttt{source\_stream} operator that takes an unbounded stream of packets as they arrive from an IP port (defined in a variable \texttt{shopping} in a Rust prelude to the Hydroflow program) and passes them to the first input (input \texttt{[0]}) of a subgraph called \texttt{lookup\_class}. Line~2 begins with a \texttt{source\_iter} operator that iterates once through a iterable Rust collection (defined in the variable \texttt{client\_class}) in a Rust prelude to the program) and passes the results to the second input (input \texttt{[1]}) of the \texttt{lookup\_class} subgraph. 

The remaining four lines specify \texttt{lookup\_class}. Line~3 specifies a relational equijoin on \texttt{(key, value)} pairs from the two inputs; inputs that match by key are concatenated in an output tuple of the form \texttt{(key, (value0, value1))}. Line~4 is a \texttt{map} function that takes the output of the join and reformats it to suit the next operator in Line~5. This is a SQL-style \texttt{group\_by} that accepts tuples of the form \texttt{(key, value)}---in this case the preceding \texttt{map} generates a key of the form \texttt{(client, class)} and a value \texttt{li}.The \texttt{group\_by} partitions the data by key, and per key it aggregates ("folds") the values using a pair of an initialization function (in this case a Rust \texttt{Vec::new} declaring an empty vector) and an iteration function (in this case \texttt{Vec::push} which pushes each tuple to the end of the vector). 
The result is a stream of tuples, one per distinct key. In Line~6 we have two operators that together do network transmission. The first is a \texttt{map} function to format tuples of the form \texttt{(payload, destination)}, and the second a \texttt{dest\_sink\_serde} that serializes each payload \texttt{m} via internal Rust libraries and ships each one to the Hydroflow node at the destination \texttt{out\_addr} (a Rust variable defined in the prelude).

\begin{figure*}
\begin{minipage}{.45\textwidth}
\begin{lstlisting}[language=rust]
source_stream(shopping_bp) -> [0]lookup_class;
source_iter(client_class) -> [1]lookup_class;
lookup_class = join() 
  -> map(|(client, (li, class))| ((client, class), li))
  -> group_by(bp_bot, bp_merge)
  -> map(|m| (m, out_addr)) -> dest_sink_serde(out);
\end{lstlisting}
\end{minipage}
\begin{minipage}{.55\textwidth}
\centering
    \includegraphics[height=1in]{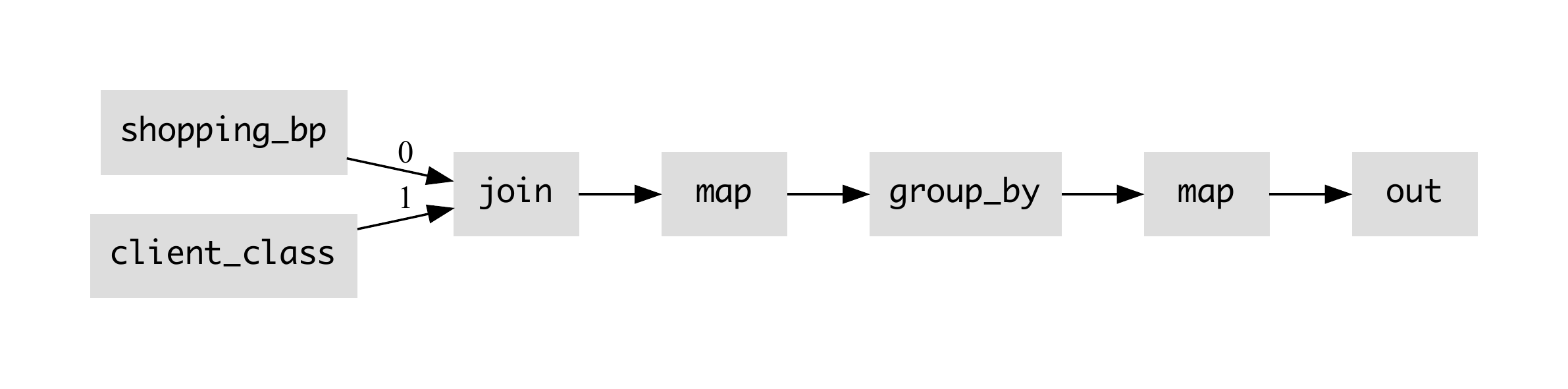}
\end{minipage}
    \caption{Bounded Prefix Lattice.}
    \label{fig:prefix}
\end{figure*}


More intuitively, this flow iterates through customer requests via the \texttt{source\_stream} operator. It then does a join with a stored \texttt{client\_class} table to look up a unique \texttt{ClientClass} tag for each client via the \texttt{join} operator, and loads the tagged shopping requests into the stateful \texttt{group\_by} operator. The \texttt{group\_by} operator is initialized with an empty vector (generated by \texttt{Vec::new}) which it accumulates by pushing each \texttt{LineItem} that arrives to the end of the list. The result is tagged via \texttt{map} with a (externally-provided) destination address \texttt{out\_addr} and sent over the network in serialized form by \texttt{dest\_sink\_serde}.

The shopping stream grows monotonically \emph{without bound}. This means that the
\texttt{group\_by} operator is never able to assemble a "final" answer for a group; even if we hacked it to ``time out'', at best it could output a ``string prefix'' (lower bound) of future answers. If we were to allow the \texttt{group\_by} to pass values to the network, then the external agent at \texttt{out\_addr} could see non-deterministic prefixes of the shopping cart. If we replicated this code to another node in the system, it might make different choices about which \texttt{group\_by} values to send out on the network, leading to inconsistency.

For correctness, then, the \texttt{group\_by} in this case can only output final results for each client ``at the end of time''---i.e. when some operational semantics of the system determines that the stream flowing into the \texttt{group\_by} will never produce more data. 
While we could augment our program with another channel for client ``checkout'' messages, that would still not help our dataflow system understand the application semantics of when to release the \texttt{group\_by} data, because checkouts could race with orders! Instead, we'd like to capture ``end-of-stream'' explicitly in our type system so the developer can inform the \texttt{group\_by} operator how to reason formally about safe release of outputs. We address this issue in the next section. 

\section{Type Upgrades for Shopping Carts}
\label{sec:types}
In this paper, we consider two kinds of transformations: \emph{type upgrades} (this section) and \emph{local graph transformations} (next section). 

The \emph{type upgrades} we seek are all \emph{bounded join semi-lattice} types. As discussed in Section~\ref{sec:background}, if we can convert to lattice types for our operators, we can override the problems introduced by networks. But we want more than just join semi-lattices; we want \emph{bounded join semi-lattices} (henceforth ``bounded lattices'') that have a well-defined finite "top" element $\top$. This element at the top of the lattice has the property that once the operator's output reaches $\top$, it will remain $\top$ in the face of any new input. This allows the operator to output the value $\top$ at any time, without fear of future retraction\footnote{Bounded semi-lattices may be the only reasonable data types to transmit across a network. Indeed, many consistency tricks in the distributed systems literature attach lattice metadata to objects in the dataflow; TCP sequence numbers, Lamport clocks and vector clocks are three common examples.
In effect this metadata ``upgrades'' the network to use lattices, but this is not typically reflected in a type system for a compiler or debugger to reason about!}.

We now proceed to show how Hydrolysis might transform our program to ``upgrade'' the types to bounded lattices.

\subsection{Bounded Prefix Lattice (BP)}
\label{sec:bp}
In this variation, the type of the incoming data is a stream of bounded lattice points $\texttt{Stream<BoundedPrefixLattice}_S\texttt{>}$. The lattice $\texttt{BoundedPrefixLattice}_S$ is defined as follows. Given some fixed-length string $S$, the domain of the lattice is a set of pairs $(s_i, \texttt{len}(S))$ containing the unique prefix of $S$ of length $i$, and the full length of $S$. The semijoin operator $\sqcup$ of this lattice takes two prefixes and simply returns the one with the longer prefix. Note that any two elements of  $\texttt{BoundedPrefixLattice}_S$ are guaranteed to share the shorter prefix by definition of both being prefixes of the same $S$. Similarly, elements of $\texttt{BoundedPrefixLattice}_S$ and $\texttt{BoundedPrefixLattice}_T$ are from different lattices and are incomparable. 
Each BP has a finite top element $\top = (S, \texttt{len}(S))$, which is identifiable in isolation---it is the only legal element whose first component $S$ has length matching the second component!
\conor{This is because these are single-source total orders rather than partial orders, right? Same trick as in TopoloTree where we know the updates come from one place and each update increments the logical time (string length in this case).} \jmh{YEP!}

We can rewrite our program of Figure~\ref{fig:orig} to use a BP without changing its output; the result is shown in Figure~\ref{fig:prefix}, with the modified \texttt{shopping\_bp} input converting each request to a BP type corresponding to its specific shopping session, and tagged with the length of the session (i.e. the number of requests in the session). Rather than streaming individual lineitems (corresponding to ``characters'' of a ``string''), it streams vector prefixes of monotonically increasing length. 
Notice how the \texttt{group\_by} operator is now initialized with the ``bottom'' ($\bot$) of the lattice via \texttt{bp\_bot}, and uses the lattice merge operator \texttt{bp\_merge} (line~5) rather than the ad-hoc vector \texttt{push} logic of Figure~\ref{fig:orig} (line~5).
Needless to say using BPs is less efficient in time and space than our original program. But this variation allows us to produce output in bounded time, and the flow is now based entirely on monotonic lattice operators, paving the way for tolerating network edges in subsequent transformations.
\conor{May be worth connecting this to the op based vs state based CRDT/lattice conversation. Instead of sending the update (op) we send the prefix (entire lattice state).}\jmh{Will handle below once we discuss the optimization to avoid resending prefixes.}


\subsubsection{Optimizations on BPs}
If we know that an edge in our dataflow is a local edge, then we know it preserves ordering and exactly-once delivery between producer and consumer. We can 
exploit that ordering semantics to implement the BP in a more efficient manner. The result will be similar to our original program based on \texttt{Vec}s, but with guarantees to allow outputs as soon as possible. Specifically, assume we have a producing operator \texttt{P} and consuming operator \texttt{Q}, and \texttt{P} emits a stream of monotonically increasing BP points (i.e. \texttt{Vec} prefixes) across a standard (non-network) edge. We can rewrite the flow segement \texttt{P~->~Q} as \texttt{P~->~odiff ->~append(len($S$)) ->~Q} where the output of $\texttt{odiff}(s_j)$ is the ``ordered diff''---the suffix of items from the input that were not produced in any previous output $s_i, i < j$---and \texttt{append(len($S$))} maintains a buffer of length $\texttt{len}(S)$ to reassemble the ordered diffs back into longer and longer prefixes. 
\conor{Its not obvious to me why this protects against duplicate delivery. Should we assume exactly once delivery in addition to order preserving on this edge from P to Q?} \jmh{thanks. ambiguity fixed via comment at start of paragraph}

This rewrite preserves the BP semantics for $P$ and $Q$ while avoiding the space consumption and data copying of redundant prefixes. Stream termination is detected when the append buffer is full. For correctness, this optimization \emph{requires} that the edge between \texttt{odiff} and \texttt{append} maintains ordering and exactly-once delivery. Ordering comes ``for free'' on local edges running on a single thread with \texttt{P} and \texttt{Q}. Note that the \texttt{append} operator is not a lattice operator: it is neither associative, nor commutative, nor idempotent. Instead it relies upon the edge itself to be ``upgraded'' to an ordered, exactly-once delivery.

This optimization opens up the possibility of more optimizations to avoid or postpone reassembling the prefixes. For example, suppose that \texttt{Q} is the operator \texttt{map(|s| uppercase(s))}.  Then \texttt{Q} runs correctly on \texttt{odiff}s, and hence we can rewrite our program by ``pushing'' \texttt{Q} earlier in the stream (\texttt{$P$ -> odiff -> $Q$ -> append}) without changing semantics. This now requires that \emph{all} edges between \texttt{odiff} and \texttt{append} must be ordered and exactly-once, but it ensures that uppercasing is only done once per character. In the most felicitous case, we are able to optimize a single-node program by ``pushing'' the \texttt{odiff} operator to the beginning of the flow, and the \texttt{append} operator to the end of the flow. In such a case, the flow becomes an intuitive local, ordered stream of small individual items. 
As a separate optimization, we may be able to fuse \texttt{odiff} into $P$, or \texttt{append} into $Q$ in certain circumstances. In our original program, the `\texttt{group\_by}` was doing precisely the work of \texttt{append}, so if it was next to an \texttt{append} operator we could entirely delete the \texttt{append} without changing semantics.

These optimizations do not always apply. Moreover, optimized BPs require local edges. As an alternative, we shift attention to an alternate (isomorphic!) structure, the Sealed Set of Indexed Values, a fully lattice-based approach that works across ``downgraded'' network edges with relatively small space overheads.

\subsection{Sealed Set of Indexed Values Lattice (SSIV)}

The idea with the SSIV is to embrace the idea of ``diffs'', but allow them to be accumulated in an ACI fashion. Borrowing ideas from TCP and Conway, et al.~\cite{blooml}, we exploit two tricks simultaneously to get a bounded lattice.
To begin, we can represent a string $S$ as a set of indexed values \texttt{(value, pos)} accumulated via union (sets with union form a lattice!), where \texttt{pos} is a natural number representing a position (index) in the string. 
Having converted from a vector to a set lattice, we can use a simple trick to bound the lattice. Specifically, a producer can count the size of the set (the length of the string) while enumerating, and piggyback the size on the last element it produces to form a bound or ``seal''. Once a consumer knows the set size and has received that many distinct elements (possibly out of order!), it knows locally---without any coordination---that it has reached the top of the lattice $\top$ and no new information will be forthcoming. Physically, our representation of items in a sealed set is a triple \texttt{(pos, val, Option$<$len($S$)$>$)}, where \texttt{pos} is an index between 0 and \texttt{len($S$) - 1}, \texttt{val} is the value in position \texttt{pos}, and \texttt{len($S$)} is an optional field---if provided, it is the length of the string. 
\conor{This SSIV trick is exactly the algorithm TCP uses to enforce ordered exactly once delivery and detect seal, right? Could be worth noting this connection.} \jmh{done}
In Figure~\ref{fig:sealed_set} we reconsider our example, using a SSIV. The code is identical to that of Figure~\ref{fig:prefix}, but using SSIVs instead of BPs.

\begin{figure}
\begin{lstlisting}[language=rust]
source_stream(shopping_ssiv) -> [0]lookup_class;
source_iter(client_class) -> [1]lookup_class;
lookup_class = join() 
  -> map(|(client, (li, class))| ((client, class), li)) 
  -> group_by(ssiv_bot, ssiv_merge)
  -> map(|m| (m, out_addr)) -> dest_sink_serde(out);
\end{lstlisting}

    \caption{Sealed Set of Indexed Values. Dataflow diagram is identical to Figure~\ref{fig:prefix},
    except \texttt{ssiv} replaces \texttt{bp}.}
    \label{fig:sealed_set}
\end{figure}

\section{Local Graph Transformations}
In this section we examing some dataflow graph transformations that, in concert with our lattice-typed shopping carts, allow us to deliver a fully monotonic, lattice-based implementation of our program. This in turn enables graph transformations for safely distributing the program without any coordination.


\subsection{Push Group By Through Join}
\begin{figure*}
  \begin{minipage}{.4\textwidth}
  \begin{lstlisting}[language=rust]
  source_stream(shopping_ssiv) 
    -> group_by(ssiv_bot, ssiv_merge)
    -> [0]lookup_class;
  source_iter(client_class) -> [1]lookup_class;
  lookup_class = join() 
    -> map(|(client, (li, class))| ((client, class), li))
    -> map(|m| (m, out_addr)) 
    -> dest_sink_serde(out);
  \end{lstlisting}
  \end{minipage}
  \begin{minipage}{.6\textwidth}
  \centering
      \includegraphics[height=1in]{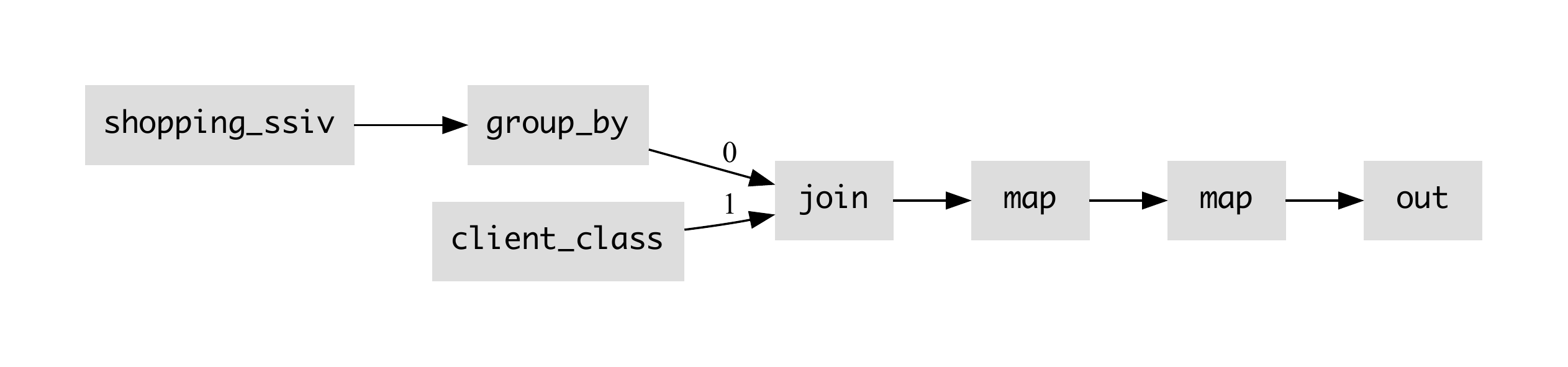}
  \end{minipage}
      \caption{Push Group By Through Join}
      \label{fig:pushgbjoin}
  \end{figure*}
In Figures~\ref{fig:prefix} and~\ref{fig:sealed_set}, we have a chain of operators \texttt{join -> map -> group\_by}, where the join finds matches based 
on \texttt{client}, the map simply rearranges the data into \texttt{(key, val)} pairs to conform to the \texttt{join} API, and the \texttt{group\_by} partitions on the pair \texttt{(client, class)}. As mentioned previously, there is one unique \texttt{class} per \texttt{client}; that is, we have a functional dependency $\mbox{\texttt{client}} \rightarrow \mbox{\texttt{class}}$. That means that the groups are partitioned uniquely by \texttt{client} alone; the \texttt{class} is simply a deterministic function of the \texttt{client}. This presents an opportunity for a classic query optimization that pushes the \texttt{group\_by} through the join (e.g.~\cite{chaudhuri1994including,eagerlazyagg}). The resulting program is shown in Figure~\ref{fig:pushgbjoin}. 

This transformation can improve performance significantly, because we now look up the client's \texttt{ClientClass} once per \emph{sealed cart} (after the \texttt{group\_by}) rather than once per lineitem request. This was the intended goal of this optimization in the early literature. Perhaps more interesting for this paper, pushing the \texttt{group\_by} down before the join means that lineitems need not be stored on the same node as the \texttt{client\_class} table, as we will discuss next.

\subsection{Decoupling Across a Network}
\label{sec:decoupling}
\begin{figure*}
  \begin{lstlisting}[language=rust]
  source_stream(shopping_ssiv) -> map(|pair| (pair, addr1))  -> dest_sink_serde(reqs_out);
  source_stream_serde(reqs_in) -> map(|((client, req), _a): ((usize, ReqSsivLattice), _)| (client, req))
    -> group_by(ssiv_bot, ssiv_merge) -> [0]lookup_class;
  source_iter(client_class) -> [1]lookup_class;
  lookup_class = join() -> map(|(client, (li, class))| ((client, class), li)) -> map(|m| (m, out_addr))
    -> dest_sink_serde(out);
  \end{lstlisting}
      \centering
      \includegraphics[width=\textwidth]{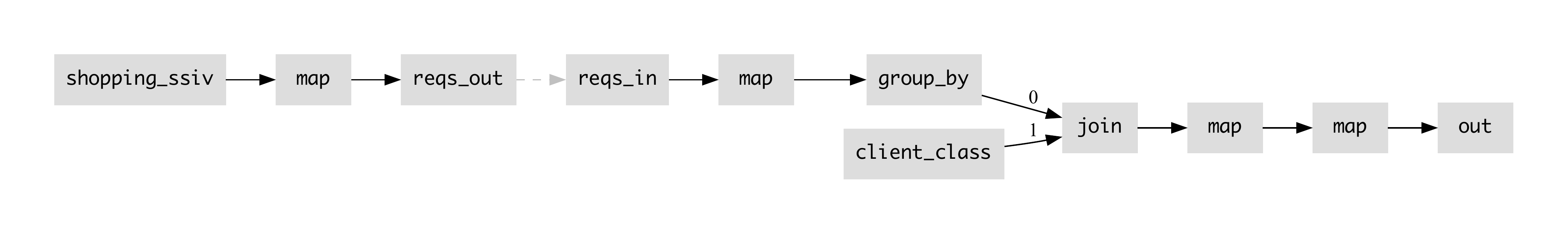}
      \caption{Decouple Across a Network: Server-Side Cart State}
      \label{fig:networked}
  \end{figure*}
  
  \begin{figure*}
  \begin{lstlisting}[language=rust]
  source_stream(shopping_ssiv) -> group_by(ssiv_bot, ssiv_merge) -> map(|pair| (pair, addr1)) -> dest_sink_serde(basic_out);
  source_stream_serde(basic_in) -> map(|((client, cart), _a): ((usize, ReqSsivLattice), _)| (client, cart))
    -> [0]lookup_class;
  source_iter(client_class) -> [1]lookup_class;
  lookup_class = join() -> map(|(client, (li, class))| ((client, class), li)) -> map(|m| (m, out_addr)) -> dest_sink_serde(out);
  \end{lstlisting}
      \centering
      \includegraphics[width=\textwidth]{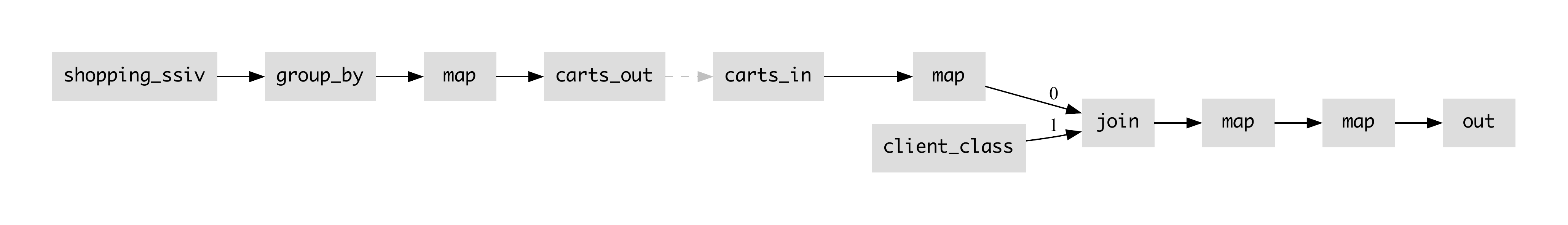}
      \caption{Decouple Across a Network: Client-Side State}
      \label{fig:networked2}
  \end{figure*}
  
Thanks to the type upgrades of Section~\ref{sec:types}, our dataflow is now fully composed of monotonic lattice operators\footnote{The only potential non-monotonic operator in our example was the \texttt{group\_by}. Relational join is a monotonic lattice \emph{morphism} over the cross-product domain of its inputs, and purely functional \texttt{map} functions are lattice morphisms as well with respect to the set of items passed into them~\cite{blooml}.}.  Note that lattices are used throughout the entire shopping cart lifecycle: not just for cart add and delete requests, but also for checkout, which is a monotone ``threshold test'' for $\top$ on the bounded semi-lattice of a session. This ``monotone checkout'' trick is the one we borrow from Conway, et al.~\cite{blooml}.
As a result of complete monotonicity, we can use network edges between any of our operators---say separating ``client'' and ``server'' components---and count on the operators to provide consistent behavior due to their ACI properties. 

In particular, notice that the \texttt{group\_by} operator maintains the state for each shopping cart. Having pushed the \texttt{group\_by} down close to the source in our previous transformation, we can now choose to ``cut the flow'' by introducing a network edge in one of two places. The first option is to put the network edges \emph{upstream} of the \texttt{group\_by}, as shown in Figure~\ref{fig:networked}. This means that clients are stateless and simply send lineitem requests to the server, which holds the cart state. This design may be useful for fault tolerance, as the server may be replicated (Section~\ref{sec:replication}) and hence be more reliable than the client. The second option is that we can put the network edge downstream of the \texttt{group\_by} as in Figure~\ref{fig:networked2}. This results in the cart state being accumulated on the client side of the network, which may be favorable for concerns of governance or privacy. Note that the \texttt{client\_class} table mixes vendor-centric information about many clients, and seems reasonable to store at the server, but the transient state of the cart is kept at the client. Hence the server only sees carts after checkout; if a client regrets adding something to their cart and subsequently deletes it before checkout, only the client will know that. This optimization choice reflects a nuanced data ownership position that sits between a fully stateless client implementation, and a \emph{local first}~\cite{localfirst} design in which no state is stored on servers. Other choices for state partitioning are possible as well via related transformation choices.


\subsection{Server Replication}
\label{sec:replication}
\begin{figure*}
  \begin{lstlisting}[language=rust]
  source_stream_serde(reqs_in) -> map(|((client, cart), _a): ((usize, ReqSsivLattice), _)| (client, cart))
      -> group_by(ssiv_bot, ssiv_merge) -> [0]lookup_class;
  source_iter(client_class) -> [1]lookup_class;
  lookup_class = join() -> map(|(client, (li, class))| ((client, class), li) ) -> tee();
  lookup_class[clients] -> all_in;
  lookup_class[broadcast] -> [0]broadcast;
  source_stream(server_addrs) -> [1]broadcast;
  broadcast = cross_join() -> dest_sink_serde(broadcast_out);
  source_stream_serde(broadcast_in) -> map(|(m, _a): (((usize, ClientClass), ReqSsivLattice), _)| m) -> all_in;
  all_in = merge() -> group_by(ssiv_bot, ssiv_merge) -> unique()
    -> map(|m| (m, out_addr)) -> dest_sink_serde(out);
  \end{lstlisting}
      \centering
      \includegraphics[width=\textwidth]{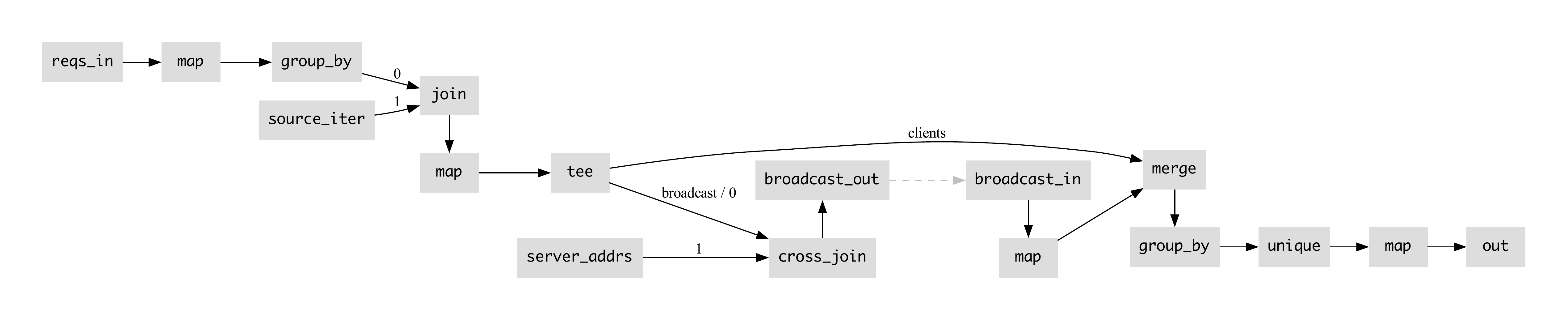}
      \caption{Replicated Server with Broadcast.}
      \label{fig:replicated}
  \end{figure*}
Another advantage of our type ``upgrade'' to lattices is that we can replicate our stateful component for fault tolerance and/or geo-locality, and have the various replicas broadcast updates amongst each other to reach eventual consistency, outputting results whenever all information becomes available ($\top$). In Figure~\ref{fig:replicated} we show a replicated version of the stateful server in Figure~\ref{fig:networked}. Note that the figure omits the unmodified ``client'' logic from line 1 of Figure~\ref{fig:networked} to keep the dataflow diagram visible.

Getting from the singleton server code to the replicated server code requires the application of a number of transformations. Space prevents us from stepping through them one by one.
In brief, the transformation flow is as follows:
\begin{enumerate}
    \item We add logic to \texttt{tee} the shopping carts to a ``broadcast'' channel (lines 4-6 of Figure~\ref{fig:replicated}).
    \item We add logic to send the broadcast to all server addresses; each message is replicated for each server via a cartesian product operator (aka \texttt{cross\_join}), and then sent to each server (line~6-8).
    \item We add logic to receive the broadcast and merge the remote updates into the local flow via an additional copy of the (idempotent!) lattice merge via \texttt{group\_by} (lines 9-10) and avoid sending redundant updates via \texttt{unique}\footnote{Some readers may note that the first \texttt{group\_by} operator is now unnecessary for correctness; it offers a sub-aggregation, but the second \texttt{group\_by} could instead aggregate individual lineitem requests from clients as well as broadcasts from replicas. It is a matter of optimization whether the earlier \texttt{group\_by} should be elided; this again fits in the realm of classical query optimization, and would be explored by Hydrolysis in a full implementation.}.
\end{enumerate}










\section{Discussion and Future Work}
\label{sec:future}
This paper captures a snapshot of our early explorations of the potential of dataflow optimization as a vehicle for optimizing distributed programs. We are increasingly optimistic that a dataflow kernel like Hydroflow is a useful optimization target for distributed systems concerns. By incorporating the ACI properties of lattices into our type system we can reason about allowing network communication to be introduced safely into programs. As illustrated in Section~\ref{sec:types} we are beginning to see the potential for an optimizer to automatically ``upgrade'' programs to use lattice types without changing semantics. This in turn opens up opportunities for decoupling and replication of program components across networks.

In Section~\ref{sec:decoupling} we saw that two different choices of program transformations can address different objectives: in that case a tradeoff between fault tolerance on one hand, and governance/privacy on another. This suggests that the objective function and constraints for optimizing modern distributed programs may be quite a bit more varied and nuanced than classical query compilation.

This workshop paper is early, and we are eagerly pursuing a number of directions to go from these concepts and hand-optimizations to a rich, automated reality. The agenda encompasses a range of challenges, including the following.
\begin{enumerate}
    \item We need a language for our own use to formalize our rewrite rules and prove equivalence of rewritten program fragments.
    \item We need to register a large number of transformation rules. This likely needs to include many classical examples from relational database query optimization, functional programming and stream query processing. In addition, we expect to trip across new optimizations that address issues with cloud deployments.
    \item We need a way for programmers to define multiple objectives they want to optimize, and to express constraints on the optimization space, e.g. for fault tolerance or governance. In the vision paper for Hydro~\cite{hydro} we highlight fault tolerance as a programming ``aspect'' that developers should be able to specify independent of their program's intended semantics. That vision requires further work, but some seeds are apparent even in our simple example here---namely the ability to consistently replicate components. Simultaneously maintaining fault tolerance constraints and performance objectives is an interesting challenge for an optimizer like Hydrolysis.
    \item We need a transformation-based optimizer that can ingest our rules, objective functions and constraints, and efficiently search the space of equivalent programs to minimize the objective function. We are enthusiastic that recent work on e-graphs could offer an efficient vehicle for our work.
\end{enumerate}
We are optimistic that open-source tools like Egg~\cite{egg} can make it possible for us to address these challenges relatively quickly. The Hydro stack itself is also open source, and we welcome additional research and development efforts!

\begin{acks}
This work was supported by unrestricted gifts from Amazon Web Services, Ant Group, Ericsson, Futurewei, Google, Intel, Meta, Microsoft, Scotiabank, VMware, Meta, Astronomer, IBM, Intel, Lacework, Mohamed Bin Zayed University of Artificial Intelligence, Nexla, Samsung SDS, Uber and NSF CISE Expeditions Award CCF-1730628. Hellerstein's work was done while on partial leave at Sutter Hill Ventures.  This work arose from many discussions with Hydro teammates Tiemo Bang, Alvin Cheung, David Chu, Natacha Crooks, Chris Douglas, Justin Jaffray, Lucky Katahanas, Chris Liu, Rithvik Panchapakesan, and Kaushik Shivakumar.
\end{acks}

\bibliographystyle{ACM-Reference-Format}
\bibliography{the-bib}










\end{document}